\documentstyle[aps,preprint]{revtex}

\begin{document}
\draft
\tightenlines
\preprint{SNUTP-98-076}
\title{Factorization and q-Deformed Algebra of Quantum Anharmonic Oscillator}

\author{Dongsu Bak, $^{1,}$\footnote{Electronic address:
dsbak@mach.uos.ac.kr}
Sang Pyo Kim,$^{2,}$\footnote{Electronic address: sangkim@knusun1.kunsan.ac.kr}
Sung Ku Kim, $^{3,}$\footnote{Electronic address: skkim@theory.ewha.ac.kr}
Kwang-Sup Soh, $^{4,}$\footnote{Electronic address: kssoh@phya.snu.ac.kr}
and Jae Hyung Yee $^{5,}$\footnote{Electronic address: jhyee@phya.yonsei.ac.kr}}

\address{$^1$ Department of Physics,
University of Seoul,
Seoul 130-743, Korea \\
$^2$ Department of Physics,
Kunsan National University,
Kunsan 573-701, Korea \\
$^3$ Department of Physics and Research Institute for Basic Sciences,
Ewha Womans University,
Seoul 120-750, Korea\\
$^4$ Department of Physics Education,
Seoul National University,
Seoul 151-742, Korea\\
$^5$ Department of Physics and Institute for Mathematical Sciences,
Yonsei University,
Seoul 120-749, Korea}

\date{\today}

\maketitle
\begin{abstract}
We have studied the underlying algebraic structure of the 
anharmonic oscillator by using the 
variational perturbation theory. To the first order of the variational perturbation, 
the Hamiltonian is found to be factorized into a supersymmetric form in terms of 
the annihilation and creation operators, which satisfy a $q$-deformed algebra. 
This algebraic structure is used to construct all the eigenstates of the Hamiltonian.
\end{abstract}

Quantum anharmonic oscillator has been frequently studied as a toy model 
for developing various approximation methods in quantum mechanics and quantum field theory 
\cite{dineykhan,justin,cea,lee,bak1}. Recently it has been used to develop various approaches to the 
variational perturbation theory \cite{cea,lee}, which enables one to compute the order by order 
correction terms to the well known variational approximation. More recently the model has been 
utilized to establish the Liouville-Neumann approach to the variational perturbation theory \cite{bak1}, 
where one constructs the annihilation and creation operators as perturbation series in the coupling 
constant whose zeroth order terms constitute those of the Gaussian approximation.

However, the underlying algebraic structure of the anharmonic oscillator for its own sake has rarely been studied.
In ref.\cite{bak1}, we have shown that to the first order of the variational perturbation the Hamiltonian 
is factorized as in the case of the simple harmonic oscillator, while the annihilation and creation 
operators satisfy the $q$-deformed algebra rather than the usual commutation relations. 
This is an interesting algebraic structure of the theory which may enable one to obtain more information 
on the theory. The connection between the $q$-deformed algebra and the quasi-exactly solvability has been 
found for certain type of potentials \cite{bonatsos}, and the possibility of a $q$-deformed quartic 
oscillator has also been suggested from the study of the energy spectra obtained by the standard 
perturbation method \cite{chung}.

It is the purpose of this letter to study the algebraic structure of the anharmonic oscillator to the 
first-order variational perturbation, and to utilize this structure to obtain the general energy eigenstates 
of the system. It is the $q$-deformed algebraic structure of the theory that enables us to find the  
$q$-deformed Fock space \cite{biedenharn}.

We now consider the anharmonic oscillator described by the Hamiltonian,
\begin{equation}
\hat{H} = \frac{1}{2}{\hat{p}}^2 + \frac{1}{2} \omega^2 {\hat{x}}^2
+ \frac{1}{4} \lambda {\hat{x}}^4,
\label{anhar}
\end{equation}
where the mass is scaled to unity for simplicity.
In the variational Gaussian approximation one searches for a simple harmonic oscillator whose energy eigenstates 
minimize the expectation value of the Hamiltonian (\ref{anhar}). For this purpose, we introduce a set of  
operators, $\hat{a}$ and $\hat{a}^{\dagger}$, as linear functions of the dynamical variables $\hat{x}$ and $\hat{p}$:
\begin{eqnarray}
\hat{a} &=& \sqrt{\frac{\Omega_G}{2 \hbar}} \hat{x}
+ i \frac{1}{\sqrt{2 \Omega_G \hbar}} \hat{p},
\nonumber\\
\hat{a}^{\dagger} &=& \sqrt{\frac{\Omega_G}{2 \hbar}} \hat{x}
- i \frac{1}{\sqrt{2 \Omega_G \hbar}} \hat{p},
\label{ann-cre-operators}
\end{eqnarray}
which are the annihilation and creation operators for the simple harmonic oscillator described by the 
Hamiltonian,  
\begin{equation}
\hat{H}_G = \frac{1}{2} {\hat{p}}^2 + \frac{1}{2} \Omega^2_G {\hat{x}}^2.
\label{har}
\end{equation}
In terms of the annihilation and creation operators (\ref{ann-cre-operators}), the Hamiltonian (\ref{anhar}) is 
represented as  
\begin{eqnarray}
\hat{H} &=& \frac{\hbar}{2} \Bigl(\frac{\omega^2}{\Omega_G}
+ \Omega_G + \frac{3 \lambda \hbar}{2 \Omega_G^2} \Bigr) \Bigl( \hat{a}^{\dagger} \hat{a} + \frac{1}{2}\Bigr)
- \frac{3 \lambda \hbar}{16 \Omega_G^2}
\nonumber \\
&+& \frac{\hbar}{4} \Bigl(\frac{\omega^2}{\Omega_G}
- \Omega_G + \frac{3 \lambda \hbar}{2 \Omega_G^2} \Bigr) \Bigl( {\hat{a}}^2 + \hat{a}^{\dagger 2} \Bigr) 
+ \frac{\lambda {\hbar}^2 }{16 \Omega_G^2 } 
\sum_{k = 0}^{4} {4 \choose k} \hat{a}^{\dagger k} \hat{a}^{4-k}.
\label{oscillator1}
\end{eqnarray}
In the variational Gaussian approximation one evaluates the expectation value of the Hamiltonian 
(\ref{oscillator1}) with respect to the state annihilated by the annihilation operator $\hat{a}$: 
\begin{equation}
\hat{a} \vert 0 \rangle_{[0]} = 0,
\label{vacuum1}
\end{equation}
and minimizes the expectation value, which leads to the gap equation,
\begin{equation}
\Omega_G^2 = \omega^2 + \frac{3 \lambda \hbar}{2 \Omega_G}.
\label{freq}
\end{equation}
The gap equation (\ref{freq}) completely determines the operator $\hat{a}$ and $\hat{a}^{\dagger}$ of 
(\ref{ann-cre-operators}), which gives the Gaussian approximation of the ground state of the system through 
Eq. (\ref{vacuum1}).

By choosing the frequency (\ref{freq}), the Hamiltonian of the anharmonic oscillator  
(\ref{oscillator1}) can now be expressed as, 
\begin{equation}
\hat{H} = \frac{\hbar}{2} \Omega_G
\Bigl( \hat{a}^{\dagger} \hat{a} + \hat{a} \hat{a}^{\dagger}
\Bigr) - \frac{3 \lambda {\hbar}^2 }{16 \Omega_G^2}
 + \frac{\lambda {\hbar}^2 }{16 \Omega_G^2}
\sum_{k = 0}^{4} {4 \choose k} \hat{a}^{\dagger k} \hat{a}^{4-k}.
\label{anhar rep}
\end{equation}
It is convenient to rewrite the Hamiltonian (\ref{anhar rep}) as,
\begin{equation}
\hat{H} = \hbar \Biggl\{ \frac{\Omega_{[1]}}{2}
\Bigl( \hat{a}^{\dagger} \hat{a} + \hat{a} \hat{a}^{\dagger}
\Bigr) + \frac{3 \lambda \hbar}{8 \Omega_G^2}
\hat{a}^{\dagger} \hat{a}
 + \frac{\lambda \hbar}{16 \Omega_G^2}
\sum_{k = 0}^{4} {4 \choose k} \hat{a}^{\dagger k} \hat{a}^{4-k}
\Biggr\},
\label{rep 2}
\end{equation}
where
\begin{equation}
\Omega_{[1]} = \Omega_G - \frac{3 \lambda \hbar}{8 \Omega_G^2},
\end{equation}
and we have used the relation, $\bigl{[} \hat{a}, \hat{a}^{\dagger} \bigr{]} = 
 \hat{a} \hat{a}^{\dagger} + \hat{a}^{\dagger} \hat{a} - 2 \hat{a}^{\dagger} \hat{a} $.
This representation can be used to construct the annihilation operator of the anharmonic oscillator as a 
perturbation series in the coupling constant, which leads to the Liouville-Neumann approach to the 
variational perturbation theory \cite{bak1}.

We now turn to the main issue of this paper: the factorization of the Hamiltonian. The solvability of the 
simple harmonic oscillator (\ref{har}) is rooted in the fact that the Hamiltonian is factorized into a 
supersymmetric form,
\begin{equation}
\hat{H}_G = \frac{\hbar}{2} \Omega_G
\Bigl( \hat{a}^{\dagger} \hat{a} +
\hat{a} \hat{a}^{\dagger} \Bigr),
\label{supersymmetric form1}
\end{equation}
and the fact that the annihilation and creation operators satisfy the standard commutation relation,
\begin{equation}
[ \hat{a}, \hat{a}^{\dagger}] = 1.
\label{commutation relation}
\end{equation}

We now ask whether the same kind of factorization occurs in the case of the anharmonic oscillator 
(\ref{anhar}):
\begin{equation}
\hat{H} = \frac{\hbar}{2}  \Omega \Bigl( \hat{A}^{\dagger} \hat{A} +
\hat{A}\hat{A}^{\dagger} \Bigr),
\label{fact}
\end{equation}
in a perturbative sense, and if it does, what kind of commutation relation does the operators
$\hat{A}$ and $\hat{A}^{\dagger}$ satisfy. The recent result of ref.\cite{bak1} suggests that one 
can find the annihilation and creation operators $\hat{A}$ and $\hat{A}^{\dagger}$ for the anharmonic 
oscillator and their commutation relation to each order of the variational perturbation 
in $\lambda$. We have found that the operators 
$\hat{A}$ and $\hat{A}^{\dagger}$ are given by, to the order $\lambda \hbar$,
\begin{eqnarray}
\hat{A}_{[1]} &=& \hat{a} + (\lambda \hbar) \sum_{k = 0}^{3}
c_k \hat{a}^{\dagger (3-k)} \hat{a}^{k}, 
\nonumber \\
\hat{A}^{\dagger}_{[1]} &=& {\Bigl( \hat{A}_{[1]} \Bigr)}^{\dagger} ,
\end{eqnarray}
where $c_k$ are constants to be determined \cite{bak1}.
The requirement that the Hamiltonian (\ref{rep 2}) be of the factorized form (\ref{fact}) determines 
the frequency $\Omega$ and the constants $c_k$'s to this order of the variational perturbation:
\begin{equation}
\Omega = \Omega_{[1]} = \Omega_G - \frac{3 \lambda \hbar}{8 \Omega_G^2 },
\label{Omega equation1}
\end{equation}
and
\begin{equation}
c_0 = \frac{1}{16 \Omega_G^3}, ~~
c_1 = \frac{6}{16 \Omega_G^3}, ~~
c_2 = \frac{3}{16 \Omega_G^3},~~
c_3 = - \frac{2}{16 \Omega_G^3}.
\label{cons 1}
\end{equation}
We thus find that the Hamiltonian of the anharmonic oscillator factorizes as, 
\begin{equation}
\hat{H} = \frac{\hbar}{2} \Omega_{[1]} \Bigl(
\hat{A}^{\dagger}_{[1]} \hat{A}_{[1]} +
\hat{A}_{[1]}\hat{A}^{\dagger}_{[1]} \Bigr) + {\cal O} (\lambda^2),
\label{fact 2}
\end{equation}
and that the annihilation and creation operators $\hat{A}$ and $\hat{A}^{\dagger}$ satisfy the commutation 
relation, 
\begin{equation}
[\hat{A}_{[1]}, \hat{A}_{[1]}^{\dagger}] = 1
+ \frac{3 \lambda \hbar}{4 \Omega_G^3}
\hat{A}_{[1]}^{\dagger} \hat{A}_{[1]},
\label{com}
\end{equation}
to this order of the variational perturbation.

Note that the commutation relation (\ref{com}) defines a $q$-deformed algebra in the form \cite{kulish},
\begin{equation}
\hat{A}_{[1]} 
- q^2_{[1]} \hat{A}^{\dagger}_{[1]} \hat{A}_{[1]} = 1,
\end{equation}
with the deformation parameter,
\begin{equation}
q^2_{[1]} = 1 + \frac{3 \lambda \hbar}{4 \Omega_G^3}.
\label{q2-equation1}
\end{equation}
This fact can be used to find all the eigenstates of the anharmonic oscillator (\ref{anhar}) by using 
the Fock space structure of a $q$-deformed oscillator \cite{biedenharn}.
The vacuum state is defined by,
\begin{equation}
\hat{A}_{[1]} \vert 0 \rangle_{[1]} = 0,
\end{equation}
and the number states by
\begin{equation}
\vert n \rangle_{[1]} = \frac{1}{\sqrt{[n]!}} \Bigl( \hat{A}^{\dagger}_{[1]} \Bigr)^n \vert 0 \rangle_{[1]},
\end{equation}
where
\begin{equation}
[n] \equiv \frac{q^{2n}_{[1]} - 1}{q^2_{[1]} -1}.
\label{n-equation1}
\end{equation}
These number states are the eigenstates of the following operators:
\begin{eqnarray}
\hat{A}^{\dagger}_{[1]} \hat{A}_{[1]}
\vert n \rangle_{[1]} &=& [n] \vert n \rangle_{[1]},  \nonumber \\
\hat{A}_{[1]} \hat{A}^{\dagger}_{[1]}
\vert n \rangle_{[1]} &=& [n+1] \vert n \rangle_{[1]}.
\end{eqnarray}
We thus find that the number states $\vert n \rangle_{[1]}$ are the eigenstates of the Hamiltonian 
corresponding to the eigenvalue,
\begin{equation}
E_{[1] n} = \frac{\hbar}{2} \Omega_{[1]}
\Bigl([n] + [n+1] \Bigr),
\label{energy eigenvalue1}
\end{equation}
to the  first order of the variational perturbation.
Using Eqs. (\ref{q2-equation1}), (\ref{n-equation1}) and (\ref{energy eigenvalue1}), we finally obtain the energy eigenvalues 
of the anharmonic oscillator, to the order $\lambda \hbar $:
\begin{equation}
E_{[1] n} = \frac{\hbar}{2} \Omega_{[1]} \Bigl(
2n + 1 + \frac{3 \lambda \hbar}{4 \Omega_G^3 } n^2  \Bigr),
\end{equation}
which agrees with those obtained by You, et al. \cite{cea}.

In summary, we have studied the algebriac structure of the anharmonic oscillator to the first-order 
in the variational perturbation theory. It has been found that the Hamiltonian is factorized into a 
supersymmetric form in terms of the annihilation and creation operators, which are expanded around those 
of the Gaussian variational approximation. It has also been shown that the anharmonic oscillator 
has the algebra of a $q$-deformed oscillator with $SU_q (1,1)$ symmetry, which was used to construct 
all the eigenstates of the Hamiltonian to this order of the variational perturbation. Further study is needed 
to see if this interesting algebraic structure is also respected by the higher order correction terms.

\vspace{.5 in}
This work was supported by the Basic Science Research Institute Program,
Korea Ministry of Education under Project No. BSRI-98-2418, BSRI-98-2425,
BSRI-98-2427, and by the Center for Theoretical Physics, Seoul National University.
SPK was also supported by the Non-Directed Research Fund, the
Korea Research Foundation, 1997, and JHY  by the Korea Science and
Engineering Foundation under Project No. 97-07-02-02-01-3.

\end{document}